\documentclass[12pt]{article}

\usepackage{graphicx}
\usepackage{amssymb}
\usepackage{amsmath}
\usepackage{amsthm}
\usepackage{subfigure}
\usepackage{setspace}
\usepackage{natbib}
\usepackage[normal,normal,bf,up]{caption2}

\allowdisplaybreaks

\newtheorem{theorem}{Theorem}

\newtheorem{lem}{Lemma}

\begin{document}
\begin{titlepage}

%\hspace{-0.5in}\begin{large}Title:\\ \hspace{-0.5in}A fresh look at maximum parsimony\end{large}
\hspace{-0.5in}\begin{large}Title:\\ \hspace{-0.5in}Parsimony via consensus\end{large}

\hspace{-0.5in}Trevor C. Bruen and David Bryant

\vfill 
%\noindent \hspace*{-0.5in}$^*$ \emph{Department of Mathematics, Berkeley University, California}\\
%\hspace*{-0.5in}$^{\dag}$ \emph{Department of Mathematics, University of Auckland, New Zealand }\\

\end{titlepage}

\pagebreak
\begin{center}
\bigskip

\begin{large}{Running Head: Parsimony via consensus}\end{large}

\vspace{1.0 in}
\bigskip


\vspace{0.5 in}

{Corresponding Author:  David Bryant 

%    Department of Mathematics, University of Auckland, 
%    
%    Auckland 1142, New Zealand 
%     
%     E-mail: d.bryant@auckland.ac.nz
%
%   Phone: +64 9 3737599
%   
%   Fax: +64 9 3737457
        
}
\end{center}
\pagebreak

\bibliographystyle{sysbio}

\begin{abstract}
The parsimony score of a character on a tree equals the number of state changes required to fit that character onto the tree. We show that for unordered, reversible characters this score equals the number of tree rearrangements required to fit the tree onto the character. We discuss implications of this connection for the debate over the use of consensus trees or total evidence, and show how it provides a link between incongruence of characters and recombination.
\end{abstract}

\subsection*{Introduction}

The (Fitch) parsimony length of a character on a tree equals the minimum number of state changes (substitutions) required to fit the character onto a tree \citep{Fitch1971}. We turn this definition on its head and show how the parsimony length of a character equals the minimum number of changes in the {\em tree} required to fit the tree onto the {\em character}. This may be a back-to-front way to look at parsimony, but it is also a useful one. We detail two applications of the result.

The first application is that this reformulation of parsimony provides a closer link between parsimony based analysis and supertree methods. We demonstrate that the maximum parsimony tree can be viewed as a type of median consensus tree, where the median is computed with respect to the {\em SPR distance} (see below). As well, the result shows how to conduct a parsimony based analysis not just on characters but on trees,  without having to recode the trees as binary character matrices. This opens the way to a hybrid between the consensus approach and the total evidence approach, where the data is a mix characters, trees, and subtrees.

The second application of our observation on parsimony is to the analysis of pairs of characters. We show that the score of the maximum parsimony tree for two characters is a simple function of the smallest number of recombinations required to explain the incongruence between the characters without homoplasy. This result provides the basis of a highly efficient test for recombination \citep{Bruen2006a}.

Here and throughout the paper we assume that all phylogenetic trees are fully resolved (bifurcating) and that by `parsimony' we refer to Fitch parsimony, where the character states are unordered and reversible. Some of the results presented here can be extended to other forms of parsimony, and possibly to incompletely resolved trees \citep{Bruen2006c}, lie beyond the scope of this paper.

Note that in this paper we are dealing with {\em unrooted} SPR rearrangements, which are those used in tree searches.  There is a related, but distinct, concept of {\em rooted} SPR rearrangements, where the rearrangements are restricted to obey a type of temporal constraint \cite{Song2003}. It is this latter class of rooted SPR rearrangments that are used to model lateral gene transfers and recombination. It would be a worthwhile, but challenging, goal to investigate whether any of the results on unrooted SPR rearrangements in this paper can be extended to rooted SPR rearrangements.

\subsection*{Linking Parsimony with SPR}

\begin{figure}[htbp] %  figure placement: here, top, bottom, or page
   \centering
   \includegraphics[width=4in]{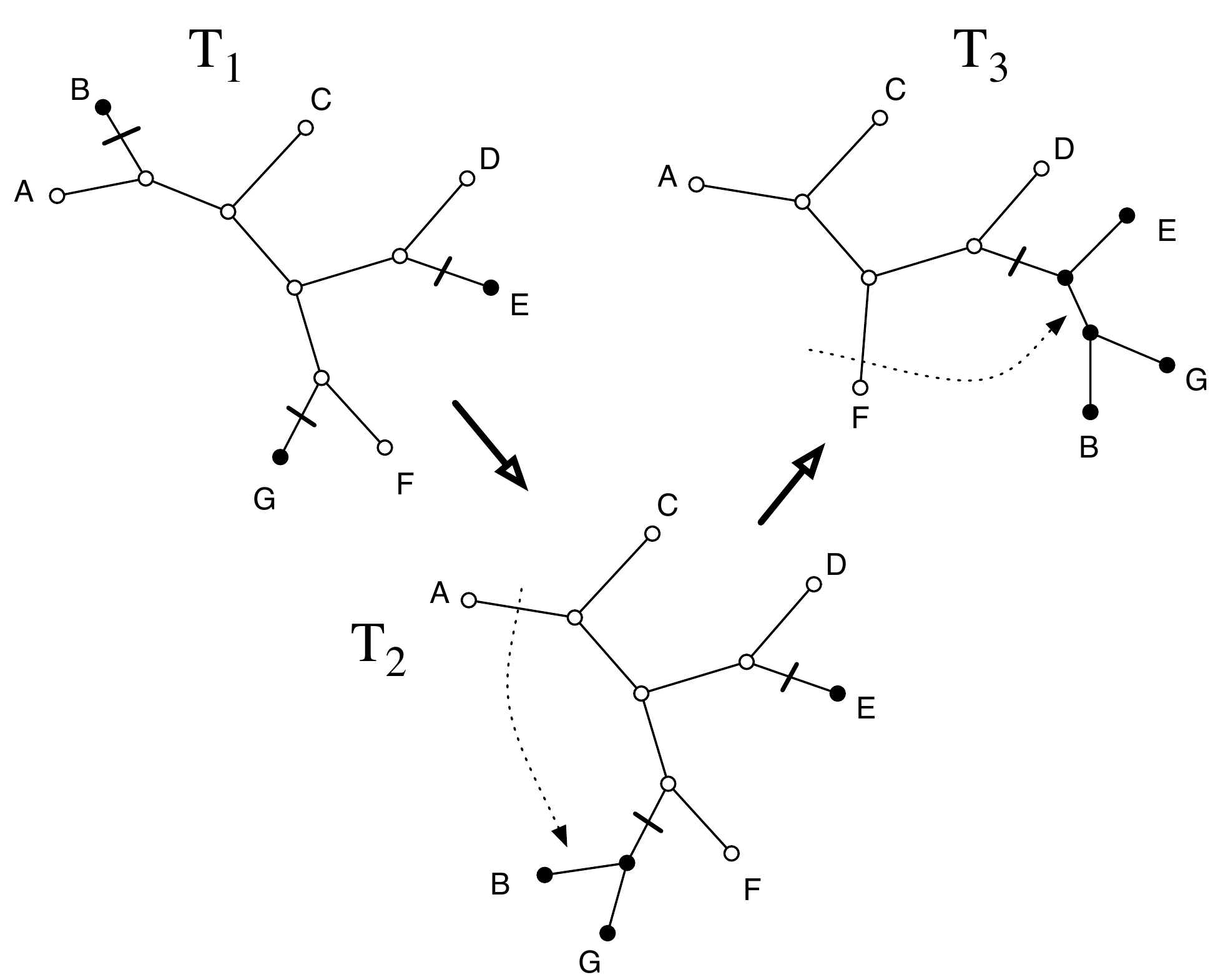} 
   \caption{Two trees, $T_1$ and $T_3$, separated by two SPR rearrangements via the intermediate tree $T_2$. A binary character of parsimony length $3$ is indicated on tree $T_1$ by the node colours. The character is compatible with a tree ($T_3$) within SPR distance two, illustrating Theorem~\ref{thm:main}.. }
   \label{fig:spr}
\end{figure}

A {\em subtree-prune and regraft} (SPR) rearrangement is an operation on phylogenetic trees whereby a subtree is removed from one part of the tree and regrafted to another part of the tree, see Figure \ref{fig:spr},  \citep{Felsenstein2004,Swofford1996}. These SPR rearrangements are widely used by tree searching software packages like PAUP \citep{Swofford1998} and Garli \citep{Zwickl2006}.  The {\em SPR distance} between two trees can be defined as the minimal number of SPR rearrangements required to transform one tree into the other  \citep{Hein1990,Allen2001,Goloboff2007}. For example, the two trees $T_1$ and $T_3$ in Figure~\ref{fig:spr}  can be transformed into each other using a minimum of two SPR  rearrangements, via the tree $T_2$, so their SPR distance is two.

The {\em parsimony length} of a character on a tree is the minimum number of steps required to fit that character on the tree, as computed by the algorithm of \citep{Fitch1971}. We will always assume unordered reversible characters The length of a character $X_i$ on a tree $T$ is denoted $\ell(X_i,T)$. A character with $r_i$ states therefore has parsimony length at least $(r_i-1)$, as every state not at the root has to arise at least once. A character is {\em compatible} with a tree if it requires at most $(r_i-1)$ changes on that tree \citep{Felsenstein2004}.

So far, one thinks of fitting a character onto a tree; we could just as well fit the tree onto the character. If the character and the tree are compatible then we have a perfect fit. When there is not a perfect fit we can measure how many SPR rearrangements are required to give a tree that does make a perfect fit. It turns out that this measure gives an equivalent score to parsimony length. More formally: 

\begin{theorem} \label{thm:main}
 Let $X_i$ be a character with $r_i$ states and let $T$ be a fully resolved phylogenetic tree. It takes exactly $\ell(X_i,T) - (r_i - 1)$ SPR rearrangements to transform $T$ into a tree compatible with $X_i$. The result still holds if $X_i$ has some missing states.
\end{theorem}

As an example, consider the character $X_1$ mapping taxa A,C,D,F to one and B,E,G to zero. The length of this character on tree $T_1$ of Figure~\ref{fig:spr} is three, and the number of SPR rearrangements needed to transform $T_1$ onto some tree $T_3$ compatible with  with $X_1$ is two. Note that there could be other trees compatible with $X_1$ are are further than two SPR rearrangements away: the result only gives the number of rearrangements required to obtain the {\em closest} tree.

Once stated, the theorem is not too difficult to prove. First show that performing an SPR rearrangement decreases the length by at most one step. Hence it takes at least $\ell(X_i,T) - (r_i - 1)$  SPR rearrangements to transform  $T$ into a tree compatible with the character $X_i$. Then show that this is the minimum required. A formal proof is presented in the Appendix. A restricted (binary character) version of this theorem was proved in \citep{Bryant2003}. 

The theorem captures an issue that is central to the interpretation  of incongruence: is an observed incongruence to be explained by positing homoplasy or by modifying the tree. Define the {\em SPR distance} from a tree $T$ to a character $X_i$ to be the SPR distance from $T$ to the closest tree $T'$ that is compatible with $X_i$. Theorem 1 then tells us that the SPR distance from $T$ is equal to the difference between the length $\ell(X_i,T)$ of $X_i$ on $T$ and the minimum possible length of $X_i$ on any tree.

\subsection*{Consensus trees, supertrees and parsimony}

In their insightful overview of supertree methods \cite{Thorley2003} characterise a family of supertree methods that all minimise a sum of the form
\begin{equation}
\sum_{i=1}^n d(T,t_i)=d(T,t_1)+d(T,t_2)+...+d(T,t_n).
\end{equation}
Here $t_1,t_2,\ldots,t_n$ are the input trees and $d(T,t_i)$ is a measure of the distance between the input tree $t_i$ with the supertree $T$. There are many choice for the distance measure $d$, and it need not be the case that the distance measure satisfies the symmetry condition $d(T,t_i) = d(t_i,T)$. \cite{Gordon1986} was the first  to propose this description of supertrees. Many supertree methods can be described in these terms, including  Matrix representation with parsimony (MRP) \citep{Baum1992,Ragan1992}; Minimum Flip supertrees \cite{Chen2006}; the Median Supertree \citep{Bryant1997}, Majority Rule Supertree \citep{Cotton2007} and the Average Consensus Supertree \citep{Lapointe1998}. 

Let $d_s(T,X_i)$ denote the SPR distance from $T$ to the closest fully resolved tree $T_i$ that is compatible with $X_i$. By Theorem 1, a maximum parsimony tree for $X_1,\ldots,X_m$ is one that minimises the expression
\begin{equation}
 \sum_{i=1}^m d_s(T,X_i)=d_s(T,X_1)+d_s(T,X_2)+...+d_s(T,X_m). \label{eq:maxPars}
  \end{equation}
In this way, maximum parsimony is a form of median consensus. The significance of this observation doesn't come from the fact that we can write the the parsimony score of $T$ in the form~\eqref{eq:maxPars}; it is from the close connection with SPR distances, and from the way we will now use this connection to combine different kinds of data in the same theoretical framework.

An {\em SPR median tree} for fully resolved trees $t_1,\ldots,t_n$ on the same leaf set is a tree $T$ that minimises 
\[\sum_{i=1}^n d_s(T,t_i)=d_s(T,t_1)+d_s(T,t_2)+...+d_s(T,t_n),\]
where here $d(T,t_i)$ denotes the SPR distance from $T$ to $t_i$ \citep{Hill2007}. We extend this directly to a supertree method by mimicking the situation for characters. Suppose that $t_i$ is a phylogenetic tree, not necessarily fully resolved, on a subset of the set of leaves. We say that a fully resolved tree $T$ on the full set of leaves is {\em compatible} with $t_i$ (equivalently, $T$ {\em displays} $t_i$) if we can obtain $t_i$ from $T$ by pruning off leaves and contracting edges. In this general situation, we let $d_s(T,t_i)$ denote the SPR distance from $T$ to the closest fully resolved tree $T_i$ that is compatible with $t_i$. This is equivalent to the more traditional definition whereby we first prune leaves off $T$ then compute the distance from this pruned tree to $t_i$.

Now suppose that we have {\em both} characters and trees in the input. Both types of phylogenetic data can be into an SPR median tree $T$, chosen to minimise the sum
\[\sum_{i=1}^n d_s(T,t_i) + \sum_{i=1}^m d_s(T,X_i).\]
We have, then, a way to bring together both the supertree/consensus methodology and the total evidence methodology. In the case that the data comprises only trees, the tree is a median supertree; in the case that the data comprises only character data, the tree is the maximum parsimony tree.

It is important to note the difference between this approach and the MRP method \citep{Baum1992,Ragan1992}, which could be used to combine trees and characters. In MRP, the  trees are broken down into multiple independent characters. This is a problem, since the characters encoding a tree are nowhere near independent. In contrast, the SPR median tree approach treats a tree as a single indivisible unit of information.   

There is one critical issue that has been side-stepped: computation time. At present, computational limitations make the construction of SPR median trees infeasible for all but the smallest data sets: just computing the SPR distance between two trees is an NP-hard problem \citep{Hickey2006}. In contrast, Total evidence and  MRP  approaches are possible for at least 100 taxa. However there are  now good heuristics for unrooted SPR distance \cite{Goloboff2007} and exact special case algorithms \cite{Hickey2006} that could be applied to the problem. Below we describe a lower bound method for the SPR distance that should also aid construction of these SPR median trees.

\subsection*{Parsimony on pairs of characters}

Another valuable application of Theorem 1 follows when we consider parsimony analysis of just two unordered and reversible characters. The concept of pairwise character compatibility was introduced by \cite{LeQuesne1969} (see also \cite{Felsenstein2004}). Two  {\em binary} characters with states $0$ and $1$ are \emph{incompatible} if and only if all four combinations of $00,01,10$, and $11$ are present as combination of states for the two characters \citep{LeQuesne1969}.  In a standard setting, character incompatibility is interpreted as implying that at least one of the characters has undergone convergent or recurrent mutation (homoplasy).  In other words, for every possible phylogeny describing the history of the two characters, at least one homoplasy is posited for one of the characters.  Another interpretation of incompatibility of two characters is that characters evolved without homoplasy on two different phylogenies, where the phylogenies differ by one or more SPR rearrangement \citep{Sneath1975, Hudson1985}.  

Define the \emph{total incongruence} score $i(X_1,X_2)$ for two multi-state unordered characters $X_1$ and $X_2$ (with $r_1$ and $r_2$ states respectively) as 
\begin{equation}
i(X_1,X_2) = \min_T \Big\{ \ell(X_1,T) + \ell(X_2,T) \Big\} - (r_1-1) - (r_2 - 1). \label{eq:incongruence}
\end{equation} 
This is the maximum  parsimony score of the two characters $X_1,X_2$ minus the minimum number of changes required for each character.  Equation~\eqref{eq:incongruence} generalises the incompatibility notion for two binary characters.  It is also equivalent to the incongruence length difference statistic applied to only two characters \citep{Farris1995}.  Importantly, the {total incongruence} score can be computed rapidly \citep{Bruen2006b}.  The following consequence of Theorem \ref{thm:main} strengthens the connection between incongruence and SPR rearrangements.

\begin{theorem} \label{thm:twochars}
The total incongruence score $i(X_1,X_2)$ for two characters  equals the minimum SPR distance between a tree $T_1$ and $T_2$ such that $X_1$ is compatible with $T_1$ and $X_2$ is compatible with $T_2$.
\end{theorem}
 
Although the notion of total incongruence for two characters has been considered before in the context of character selection and weighting \citep{Penny1986}, it has not been considered in the context of genealogical similarity.  Essentially, Theorem \ref{thm:twochars} shows that the total incongruence score equals the minimum possible number of SPR rearrangements that could have occurred between the phylogenetic histories for both characters, assuming that the characters have different histories with which they are each perfectly compatible.

Indeed, Theorem \ref{thm:twochars} suggests a natural way to interpret genealogical similarity between two characters, which we have used to develop a powerful test for recombination \citep{Bruen2006a}. Choosing two characters from two different genes (which have possibly different histories) gives a simple approach to identify the distinctiveness of the histories of the genes. 
 
 We can also apply Theorem~\ref{thm:twochars} to obtain a lower bound on an SPR distance between two trees. Suppose that we have two trees $T_1$ and $T_2$ and we wish to obtain a lower bound on the SPR distance $d(T_1,T_2)$ between the two trees. If we choose any character $X_1$ convex on $T_1$ and any character $X_2$ convex on $T_2$ then, by Theorem~\ref{thm:twochars}, we have that $i(X_1,X_2) \leq d(T_1,T_2)$. By carefully choosing $X_1$ and $X_2$ we can obtain tighter bounds.  One natural starting point for $X_1$ and $X_2$ is the four or five character encodings described by \citep{Semple2002,Huber2005}.

\subsection*{Discussion and extensions}

We have presented a reformulation of parsimony that is, in some way, dual to the standard definitions. Instead of measuring how well a character fits onto a tree we look at how well the tree fits onto the character. A consequence of this new perspective is that we can combine trees and character data using one general SPR framework, and we also obtain new results connecting incongruence measures and recombination. Nevertheless, it is not immediately clear how the new reformulation can be interpreted in itself.

\begin{figure}[htbp] %  figure placement: here, top, bottom, or page
   \centering
   \includegraphics[width=4in]{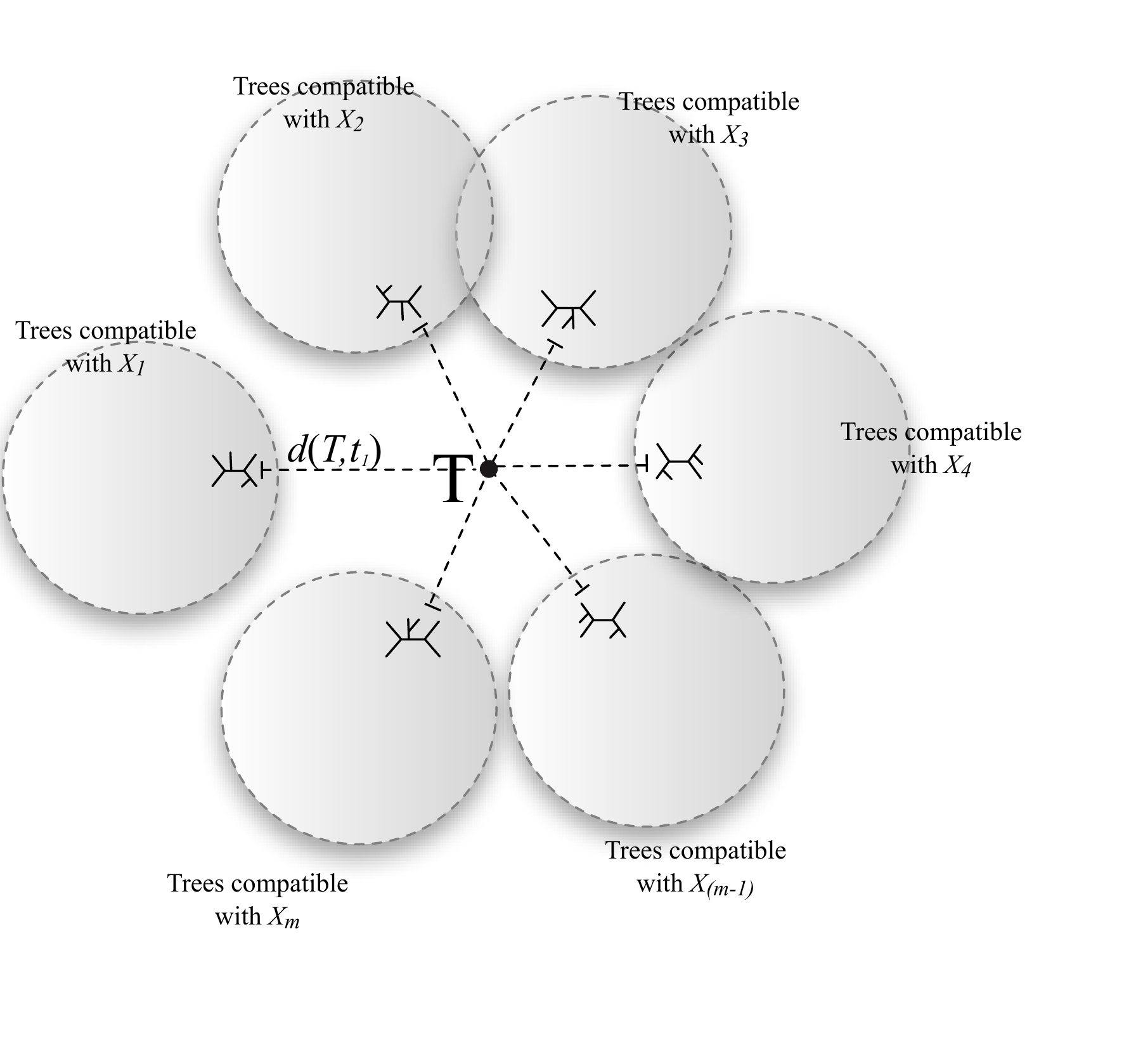} 
   \caption{Cartoon representation of parsimony in terms of tree rearrangements. Each character $X_i$ gives a `cloud' of trees containing those trees compatible with $X_i$. The maximum parsimony tree is then the tree closest to these clouds under the SPR distance.}
   \label{fig:manyClouds}
\end{figure}

One aid in this direction is to consider the information a single character, or tree, represents. Given a single character, we can imagine a {\em cloud} of trees comprising exactly those trees compatible with the character (Figure~\ref{fig:manyClouds}). If we are told that this character evolved without homoplasy, then we know that the true evolutionary tree must be contained somewhere within the cloud. However as there is only one character there is a lot of uncertainty regarding the tree, so there are a lot of trees in the clouds. Now suppose we have multiple characters, each with its own cloud.  There may not be a single tree contained in the intersection of all of these clouds. Instead, we search for a tree that is close as possible to all of the clouds. The distance from $T$ to the cloud associated to character $X_i$ is exactly $d_s(T,X_i)$, so by Theorem 1 a tree closest to all of the clouds is a maximum parsimony tree.

Each cloud represents the uncertainty around each piece of data (tree or character). \\

We note that several of the results in this article can be extended, for details. Firstly, both Theorems 1 and 2 are both valid if we replace the SPR distance with the {\em tree bisection and reconnection} (TBR) distance. In a TBR rearrangement, a subtree is removed from the tree and then reattached elsewhere in a tree, the difference with SPR being that we can reattach using any of the nodes in the subtree \citep{Allen2001,Felsenstein2004}. The TBR distance between two trees is the minimum number of TBR rearrangements required to transform one tree into the other. 

That Theorems 1 and 2 hold for the TBR distance might seem surprising, since the TBR distance between two trees is always less than, or equal to, the SPR distance between the trees. However the extension follows by a tiny change to the proof of Theorem 1, noting that a TBR move can still only reduce the parsimony score of a character by at most one. 

We have also explored extensions of the result to other distances between trees, notably the Robinson-Foulds or partition distance and the Nearest Neighbor Interchange distance, though the connections are not so clear. See \cite{Bruen2006c} for details.

\subsection*{{Acknowledgements}}
  
We would like to thank Mike Steel, Sebastien B\"{o}cker, Olaf Bininda Emonds, Pablo Golloboff, Mark Wilkinson and an anonymous referee for their valuable suggestions. This research was partially  supported by the New Zealand Marsden Fund.

   \bibliography{fullpaperbib}

\appendix

\subsection*{Appendix}
 Refer to  \citep{Semple2003} for  a detailed description of the notation. 
  
The first observation is that an TBR rearrangement of a tree increases the length of a character by at most one. As SPR rearrangements are a special case of TBR rearrangements, the same result holds for SPR.

\begin{lem} \label{wlem_a}
Let $T$ be a fully resolved phylogenetic tree and $X_i$ an unordered reversible character.  Let $T'$ be a phylogenetic tree that differs from $T$ by a single TBR rearrangement.  Then $\ell(\chi,T') \le \ell(\chi,T) +1$.  
\end{lem}
\begin{proof}
The proof of Lemma 5.1 in \citep{Bryant2004} for binary characters applies directly to the multistate case.
\end{proof}
 
 Let $d_{SPR}(T,T')$ denote the unrooted SPR distance between two phylogenetic trees $T$ and $T'$.

 \medskip
 {\bf Theorem \ref{thm:main}} {\sl  Let $X_i$ be a character with $r_i$ states and let $T$ be a fully resolved phylogenetic tree. It takes exactly $\ell(X_i,T) - (r_i - 1)$ SPR rearrangements to transform $T$ into a tree compatible with $X_i$. The result still holds if $X_i$ has some missing states.}
 
  \medskip

 \begin{proof}
Let $T'$ be a fully resolved phylogenetic tree compatible with $X_i$  for which $d_{SPR}(T,T')$ is minimized and let $m=d_{SPR}(T,T')$. Then there exists a sequence of trees $T'=T_0, ...,T_m=T$ such that every adjacent pair of trees in the sequence differ by exactly one SPR rearrangement. By Lemma \ref{wlem_a}  the existence of this sequence implies that $\ell(T,X_i) - \ell(T',X_i)  \le d_{SPR}(T,T')$ and since $X_i$ is compatible with $X_i$ we have $ \ell(T',X_i) = r_i - 1$, giving
\[\ell(T,X_i) - (r_i - 1) \le d_{SPR}(T,T').\]

For the other direction, we show that we can construct a sequence of $\ell(T,X_i) - (r_i - 1)$ SPR rearrangements that transform $T$ into a tree $T'$ compatible with $X_i$.  Firstly, if $\ell(T,X_i) - (r_i-1) = 0$, then $T$ is compatible with $X_i$ so the proof is finished.  Otherwise, let $\widehat{X_i}$ be an assignment of states to internal nodes that minimises the number of state changes (that is, a {\em minimum extension} of $X_i$).  Then since $X_i$ is not convex on $T$ there exist three vertices $u,v$ and $w$, where $\{u,v\}\in E(T)$, $v$ lies on the path from $u$ to $w$ and $\widehat{X_i}(u)=\widehat{X_i}(w)\ne \widehat{X_i}(v)$.  Perform an SPR rearrangement by removing edge $\{u,v\}$, supressing the $v$ vertex and creating a new edge $\{u,t\}$ where $t$ is a new vertex on an edge adjacent to $w$.  Furthermore, set $\widehat{X_i}(t)=\widehat{X_i}(w)$.  Then the number of edges on which a change has occurred has decreased by $1$ thereby decreasing the parsimony length by $1$.  This procedure can be repeated until the parsimony length equals $r_i - 1$, constructing the desired sequence of trees and completing the proof. 
\end{proof}

Let ${T}$ be a maximum parsimony phylogenetic tree for $X_1$ and $X_2$ and let

\medskip
{\bf Theorem \ref{thm:twochars}} {\sl The total incongruence score $i(X_1,X_2)$ for two characters  equals the minimum SPR distance between a tree $T_1$ and $T_2$ such that $X_1$ is compatible with $T_1$ and $X_2$ is compatible with $T_2$.}

\medskip

\begin{proof}
Let $T_1$ and $T_2$ be any two trees compatible with  $X_1$ and $X_2$ respectively. Then $\ell(X_1,T_1)=r_1 - 1$ and by Theorem~\ref{thm:main}, $\ell(X_2,T_1) - (r_2-1)   \leq d_{SPR}(T_1,T_2)$. We have then that 
\begin{eqnarray*}
i(X_1,X_2) &\leq& \ell(X_1,T_1) + \ell(X_2,T_1) - (r_1 - 1) - (r_2 - 1) \\
& \leq &  d_{SPR}(T_1,T_2)
\end{eqnarray*}
and so $i(X_1,X_2)$ is a lower bound for $d_{SPR}(T_1,T_2)$.

We show that this bound can be achieved. Let $T$ be a maximum parsimony tree for the pair of characters $X_1,X_2$.  By Theorem \ref{thm:main}  there exist two trees $T_1$ and $T_2$ such that $T_1$ is compatible with $X_1$, $T_2$ is compatible with $X_2$ and 
\[d_{SPR}(T_1, T) + d_{SPR}(T_2,T) = i(X_1,X_2),\]
implying that $d_{SPR}(T_1,T_2) \leq  d_{SPR}(T_1, T) + d_{SPR}(T_2,T)  \leq i(X_1,X_2)$ and hence
\[d_{SPR}(T_1,T_2) = i(X_1,X_2).\]
\end{proof}

\end{document}